\documentclass[default]{ancillary/sn-jnl_test}

\usepackage{amsmath,amssymb,amstext}
\usepackage{bookmark}
\usepackage[T1]{fontenc}
\usepackage{graphicx}
\definecolor{ForestGreen}{HTML}{2e8b21}
\definecolor{NatureRed}{HTML}{f54e42}
\definecolor{NatureBlue}{HTML}{4254f5}
\definecolor{NatureGreen}{HTML}{138f57}
\definecolor{NatureMagenta}{HTML}{6b1345}
\definecolor{NaturePurple}{HTML}{c227b0}
\definecolor{NatureCyan}{HTML}{2b807b}
\definecolor{NatureOrange}{HTML}{f5a142}
\usepackage{hyperref}
\hypersetup{
    colorlinks=true,
    linkcolor=blue,
    citecolor=blue,
    filecolor=magenta,      
    urlcolor=cyan,
    }
\usepackage{subcaption}
\usepackage{txfonts}
\usepackage{textgreek}
\usepackage{xargs}
\usepackage{xcolor}
\usepackage{xspace}
\usepackage{anyfontsize}
\usepackage{placeins}

\graphicspath{{./}{figs/}}

\jyear{2024}

\usepackage{etoolbox}
\makeatletter
\newcommand\sendemail[4]{
\edef\@tempa{mailto:#1?subject=#2&body=#3 }%
\edef\@tempb{\expandafter\html@spaces\@tempa\@empty}%
\href{\@tempb}{#4}}

\catcode\%=11
\def\html@spaces#1 #2{#1
\catcode\%=14
\makeatother

\catcode\%=11
\def\html@spaces#1 #2{#1
\catcode\%=14
\makeatother

\newcommand{\citationneeded}{\textcolor{ForestGreen}{$^{\rm citation\;needed}$}}
\let\oldtextsigma\textsigma
\renewcommand{\textsigma}{\oldtextsigma\xspace}
\let\oldAA\AA
\renewcommand{\AA}{\text{\oldAA}\xspace}
\let\oldtextdegree\textdegree
\renewcommand{\textdegree}{\oldtextdegree\xspace}
\def\w80{\ensuremath{w_{80}}\xspace}

\newcommand{\kms}{\ensuremath{\mathrm{km\,s^{-1}}}\xspace}
\newcommand{\MSun}{\ensuremath{{\rm M}_\odot}\xspace}
\newcommand{\yr}{\ensuremath{{\rm yr}}\xspace}
\newcommand{\Myr}{\ensuremath{{\rm Myr}}\xspace}
\newcommand{\Gyr}{\ensuremath{{\rm Gyr}}\xspace}
\newcommand{\peryr}{\ensuremath{{\rm yr^{-1}}}\xspace}
\newcommand{\Lsun}{\hbox{\,${\rm L}_\odot$}}
\newcommand{\mum}{\text{\textmu m}\xspace}
\newcommand{\kpc}{\text{kpc}\xspace}
\newcommand{\ZH}{\text{[Z/H]}\xspace}

\newcommandx{\lambdar}[2][1=R,2=]{\ensuremath{\lambda_{\rm {#1}}{#2}}\xspace}
\newcommand{\eps}{\ensuremath{\epsilon}\xspace}
\newcommand{\Mstar}{\ensuremath{M_\star}\xspace}
\newcommand{\Mdyn}{\ensuremath{M_\mathrm{dyn}}\xspace}
\newcommand{\re}{\ensuremath{R_\mathrm{e}}\xspace}
\newcommand{\vstar}{\ensuremath{v_\star}\xspace}
\newcommand{\vnai}{\ensuremath{v_{\NaI}}\xspace}
\newcommand{\sigmastar}{\ensuremath{\sigma_\star}\xspace}
\newcommand{\sigmaestar}{\ensuremath{\sigma_{\star,\mathrm{e}}}\xspace}
\newcommand{\vperc}[1]{\ensuremath{v_{#1}}\xspace}

\newcommand{\vesc}{\ensuremath{v_\mathrm{esc}}\xspace}
\newcommand{\nelec}{\ensuremath{n_\mathrm{e}}\xspace}
\newcommand{\Rout}{\ensuremath{R_\mathrm{out}}\xspace}
\newcommand{\vout}{\ensuremath{v_\mathrm{out}}\xspace}
\newcommandx{\Mout}[2][1=,2=]{\ensuremath{M_{\mathrm{out}{#2}}^{#1}}\xspace}
\newcommandx{\Mdotout}[2][1=,2=]{\ensuremath{\dot{M}_{\mathrm{out}{#2}}^{#1}}\xspace}

\newcommandx{\fluxdcgs}[1][1=-20]{$\times 10^{[#1]}$~erg~s$^{-1}$~cm$^{-2}$~\AA$^{-1}$\xspace}
\newcommandx{\powercgs}[1][1=44]{$\times 10^{#1}$~erg~s$^{-1}$\xspace}
\newcommand{\Av}{\ensuremath{A_V}\xspace}

\newcommand{\hda}{\ensuremath{\mathrm{H\text{\textdelta}_A}}\xspace}
\newcommand{\hga}{\ensuremath{\mathrm{H\text{\textgamma}_A}}\xspace}
\newcommand{\Halpha}{\text{H\textalpha}\xspace}
\newcommand{\Hbeta}{\text{H\textbeta}\xspace}
\newcommand{\Hgamma}{\text{H\textgamma}\xspace}
\newcommand{\Hdelta}{\text{H\textdelta}\xspace}
\newcommand{\Pabeta}{\text{Pa\textbeta}\xspace}
\newcommand{\Hepsilon}{\text{H\textepsilon}\xspace}
\newcommandx{\permittedEL}[6][1=O,2=III,3=,4=,5=,6=]{\text{{#1}\,{\sc {#2}}{#3}{#4}{#5}{#6}}\xspace}
\newcommandx{\semiforbiddenEL}[6][1=O,2=III,3=,4=,5=,6=]{\text{{#1}\,{\sc{#2}}]{#3}{#4}{#5}{#6}}\xspace}
\newcommandx{\forbiddenEL}[6][1=O,2=III,3=,4=,5=,6=]{\text{[{#1}\,{\sc{#2}}]{#3}{#4}{#5}{#6}}\xspace}

\newcommand{\CaII}{\permittedEL[Ca][ii]}
\newcommand{\OIII}{\forbiddenEL[O][iii]}
\newcommandx{\OIIIL}[1][1=5007]{\forbiddenEL[O][iii][\textlambda][#1]}
\newcommand{\OIIIall}{\forbiddenEL[O][iii][\textlambda][\textlambda][4959,][5007]}
\newcommandx{\NIL}{\forbiddenEL[N][i][\textlambda][5200]}
\newcommand{\OI}{\forbiddenEL[O][i]}
\newcommand{\OIall}{\forbiddenEL[O][i][\textlambda][\textlambda][6300,][6364]}
\newcommand{\HeI}{\permittedEL[He][i]}
\newcommand{\NaI}{\permittedEL[Na][i]}
\newcommand{\NII}{\forbiddenEL[N][ii]}
\newcommandx{\NIIL}[1][1=6583]{\forbiddenEL[N][ii][\textlambda][#1]}
\newcommand{\NIIall}{\forbiddenEL[N][ii][\textlambda][\textlambda][6548,][6583]}
\newcommand{\SII}{\forbiddenEL[S][ii]}
\newcommand{\SIIall}{\forbiddenEL[S][ii][\textlambda][\textlambda][6716,][6731]}

\newcommandx{\target}[1][1=]{\text{GS-10578{#1}}\xspace}

\newcommand{\jwst}{\textit{JWST}\xspace}
\newcommand{\hst}{\textit{HST}\xspace}
\newcommand{\ppxf}{{\sc ppxf}\xspace}

\newcommand{\Mdynvalue}{$\Mdyn = 2.0\pm0.5 \times 10^{11}$~\MSun}

\newcommand{\mstar}{\ensuremath{M_\star}\xspace}
\newcommand{\Msun}{\ensuremath{\mathrm{M}_\odot}\xspace}

\newcommand{\HeIL}[1][1=5876]{\permittedEL[He][i][\textlambda][#1]}
\newcommand{\MgIIall}{\permittedEL[Mg][ii][\textlambda][\textlambda][2796,][2803]}

\newcommand{\NaIall}{\permittedEL[Na][i][\textlambda\textlambda][5890,][5896]}
\newcommand{\CaIIall}{\permittedEL[Ca][ii][\textlambda\textlambda][3934,][3968]}

\raggedbottom

\begin{document}

\title[Low gas content in a distant quiescent galaxy]{Measurement of the gas consumption history of a massive quiescent galaxy}

\author*[1,2]{\fnm{Jan} \sur{Scholtz}}\email{js2685@cam.ac.uk}\equalcont{These authors contributed equally to this work.}
\author[1,2]{\fnm{Francesco} \sur{D'Eugenio}}
\equalcont{These authors contributed equally to this work.}
\author[1,2,3]{\fnm{Roberto} \sur{Maiolino}}
\author[4]{\fnm{Pablo~G.} \sur{P\'erez-Gonz\'alez}}
\author[5,3]{\fnm{Chiara} \sur{Circosta}}
\author[1,2]{\fnm{Sandro} \sur{Tacchella}}
\author[6]{\fnm{Christina~C.} \sur{Williams}}
\author[7]{\fnm{Stacey} \sur{Alberts}}
\author[4]{\fnm{Santiago} \sur{Arribas}}
\author[1,2]{\fnm{William~M.} \sur{Baker}}
\author[8]{\fnm{Elena} \sur{Bertola}}
\author[9]{\fnm{Andrew~J.} \sur{Bunker}}
\author[10]{\fnm{Stefano} \sur{Carniani}}
\author[11]{\fnm{Stephane} \sur{Charlot}}
\author[8]{\fnm{Giovanni} \sur{Cresci}}
\author[9]{\fnm{Gareth~C.} \sur{Jones}}
\author[12]{\fnm{Nimisha} \sur{Kumari}}
\author[4]{\fnm{Isabella} \sur{Lamperti}}
\author[1,2]{\fnm{Tobias~J.} \sur{Looser}}
\author[4]{\fnm{Bruno} \sur{Rodr\'iguez Del~Pino}}
\author[13]{\fnm{Brant} \sur{Robertson}}
\author[10]{\fnm{Eleonora} \sur{Parlanti}}
\author[4]{\fnm{Michele} \sur{Perna}}
\author[1,2]{\fnm{Hannah} \sur{\"Ubler}}
\author[10]{\fnm{Giacomo} \sur{Venturi}}
\author[1,2]{\fnm{Joris} \sur{Witstok}}

\affil[1]{Kavli Institute for Cosmology, University of Cambridge, Madingley Road, Cambridge, CB3 OHA, UK}
\affil[2]{Cavendish Laboratory - Astrophysics Group, University of Cambridge, 19 JJ Thomson Avenue, Cambridge, CB3 OHE, UK}
\affil[3]{Department of Physics and Astronomy, University College London, Gower Street, London WC1E 6BT, UK}
\affil[4]{Centro de Astrobiolog\'{\i}a (CAB), CSIC-INTA, Ctra. de Ajalvir km 4, Torrej\'on de Ardoz, E-28850, Madrid, Spain}
\affil[5]{European Space Agency (ESA), European Space Astronomy Centre (ESAC), Camino Bajo del Castillo s/n, 28692 Villanueva de la Cañada, Madrid, Spain}
\affil[6]{NSF’s National Optical-Infrared Astronomy Research Laboratory, 950 North Cherry Avenue, Tucson, AZ 85719, USA}
\affil[7]{Steward Observatory, University of Arizona, 933 N. Cherry Avenue, Tucson, AZ 85721, USA}
\affil[8]{INAF - Osservatorio Astrofisico di Arcetri, largo E. Fermi 5, 50127 Firenze, Italy}
\affil[9]{University of Oxford, Department of Physics, Denys Wilkinson Building, Keble Road, Oxford OX13RH, United Kingdom}
\affil[10]{Scuola Normale Superiore, Piazza dei Cavalieri 7, I-56126 Pisa, Italy}
\affil[11]{Sorbonne Universit\'e, UPMC-CNRS, UMR7095, Institut d'Astrophysique de Paris, F-75014 Paris, France}
\affil[12]{European Space Agency, c/o STScI, 3700 San Martin Drive, Baltimore, MD 21218, USA}
\affil[13]{Department of Astronomy and Astrophysics, University of California, Santa Cruz, 1156 High Street, Santa Cruz CA 96054, USA}

\abstract{\unboldmath
    JWST is discovering increasing numbers of quiescent galaxies 1--2 billion years after the Big Bang, whose redshift, high mass, and old stellar ages indicate that their formation and quenching were surprisingly rapid. This fast-paced evolution seems to require that feedback from AGN (active galactic nuclei) be faster and/or more efficient than previously expected. We present deep ALMA observations of cool molecular gas (the fuel for star formation) in a massive, fast-rotating, quiescent galaxy at $z=3.064$. This galaxy hosts an AGN, driving neutral-gas outflows with a mass-outflow rate of $60\pm20~\Msun$~yr$^{-1}$, and has a star-formation rate of $<5.6$~$\Msun$~yr$^{-1}$. Our data reveal this system to be a distant gas-poor galaxy confirmed with direct CO observations (molecular-gas mass $< 10^{9.1}$ \Msun; <0.8\% of its stellar mass). Combining ALMA and JWST observations, we estimate the gas-consumption history of this galaxy, showing that it evolved with net zero gas inflow, i.e., gas consumption by star formation matches the amount of gas this galaxy is missing relative to star-forming galaxies. This could arise both from preventative feedback stopping further gas inflow, which would otherwise refuel star formation or, alternatively, from fine-tuned ejective feedback matching precisely gas inflows.
}

\maketitle

Astronomers have now detected a number of massive quiescent galaxies up to 1.2 billion years after Big Bang \citep{Glazebrook17,Schreiber18,Carnall_spec,Nanayakkara24,Glazebrook2023, deGraaff24, Perez-Gonzalez24}, presenting a question of how these monsters managed to grow and die in such a short span of time. With \jwst, it is now possible to accurately constrain the star-formation histories of these massive high-redshift quiescent galaxies (e.g. \citep[]{Carnall_spec, Glazebrook2023,Park24}) even on spatially resolved scales \citep{Perez-Gonzalez24}. The short available time to form and quench these galaxies seems to require faster and/or more efficient quenching mechanisms, which brought back into vogue ejective feedback from AGN winds (e.g.,\citep{Xie24}), as well as more efficient star formation in the early Universe. However, to test whether ejective feedback is really responsible for quenching massive galaxies at high redshift, it is still necessary to observe the cool (T$\sim$ 40 K) gas supply of these galaxies, the fuel for star formation. By redshift $z=1.5\text{--}2$, i.e. 3--4~Gyr after the Big Bang, observations of quiescent galaxies show little cool, molecular gas \citep{Belli+21, Williams21,Woodrum22}. However, one of the current major questions in extra-galactic astronomy is whether the lack of molecular gas is explained either through ejective feedback, i.e. the removal of cool gas from the galaxy interstellar medium (ISM) via AGN or supernova-driven outflows or through preventative
feedback, the interruption of cosmic gas accretion accompanied by consumption of cool gas previously present in the ISM. In this work, we present deep ALMA and NIRSpec/MSA observations of \target, a massive quiescent galaxy (based on UVJ colours \citep{Barro23, DEugenio23a} at redshift $z=3.064$. This galaxy is hosting a powerful type-2 obscured AGN, detected in X-ray \citep{Luo17}, MIR \citep{Barro19} and radio \citep{Alberts20},  with an estimated bolometric luminosity of $5\times 10^{46}$ erg s$^{-1}$. Recent observations measured the stellar mass $\Mstar=1.6\pm0.2\times10^{11}~\MSun$, with 
evidence for stellar rotation, suggesting that quenching occurred without destroying the stellar disc \citep{DEugenio23a}. The detection of ionised (e.g. \OIIIall) and neutral-phase (\NaIall)
outflows with high mass loading ($\sim$60 \Msun yr$^{-1}$, with star formation of <5.6 \Msun yr$^{-1}$) makes it possible to study how efficiently AGN can remove the gas reservoir. Furthermore, both \citep{DEugenio23a, Perna23b} observe multiple satellites, Ly$\alpha$ emitters and AGN candidates within 30 kpc, suggesting a complex environment. Indeed, the system is merging with multiple low-mass satellites and is undergoing powerful, ejective feedback from its supermassive black hole.

\begin{figure}
    \centering
    \includegraphics[width=0.99\textwidth]{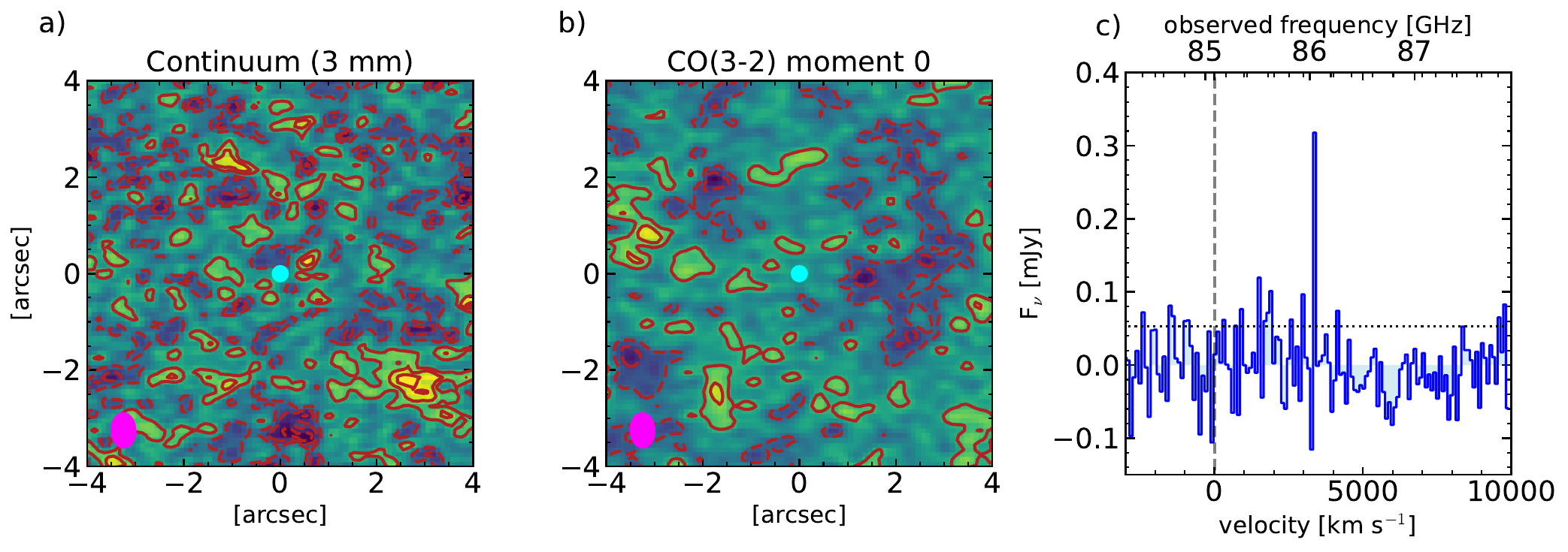}   
	\includegraphics[width=0.99\textwidth]{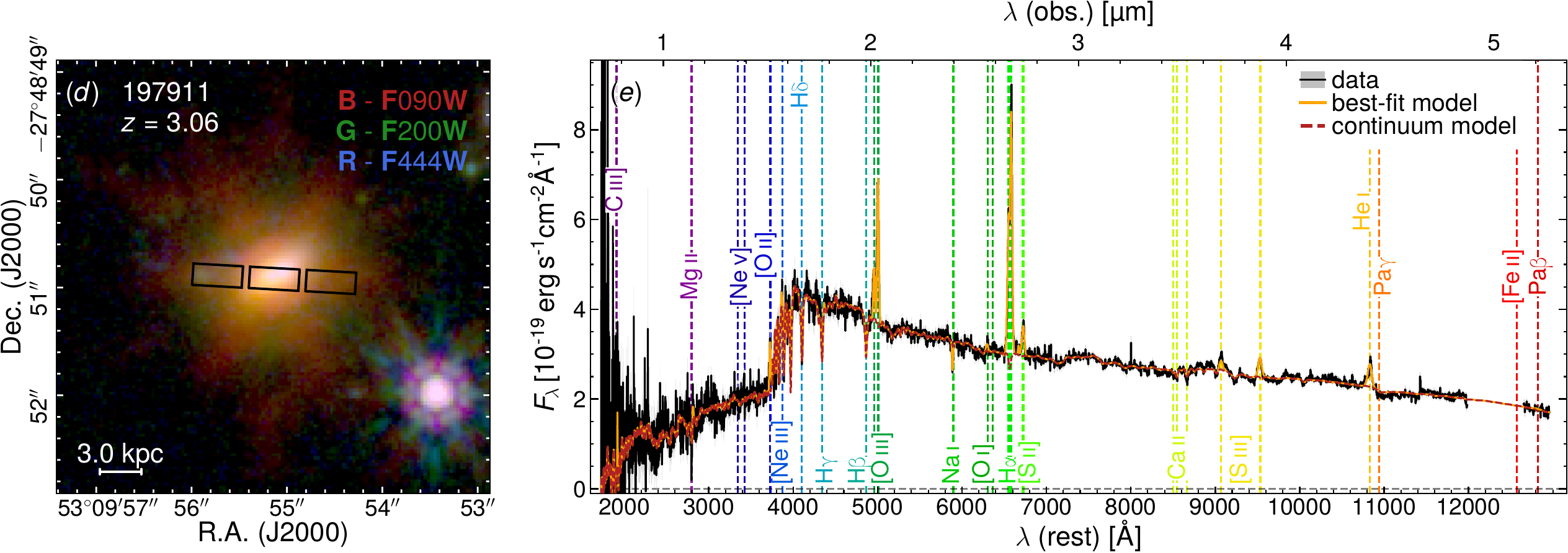}   
    \caption{Overview of the ALMA and JWST data. a) Map of the 3 mm continuum. The red contours show RMS, 2$\times$RMS and 3$\times$RMS (solid), and -RMS, -2$\times$RMS and -3$\times$RMS (dashed contours) of the displayed image. We do not see any significant detection of 3 mm continuum; b) Image of the CO(3-2) $\pm$150 km s$^{-1}$ around systematic redshift of our main target. The red contours show RMS, 2$\times$RMS and 3$\times$RMS (solid), and -RMS, -2$\times$RMS and -3$\times$RMS (dashed contours) of the displayed image. We also do not see any significant CO(3-2) emission at the location of our galaxy. In both maps, the magenta ellipse shows the ALMA beam size. In both panels a) and b) the cyan circle shows the location of our target. c) CO(3-2) spectrum extracted from the location of GS-10578 with 0.3 arcsec aperture radius. The vertical dashed line indicates the expected location of the CO(3-2) emission line;
    d) \jwst/NIRCam RGB image of GS-10578 from the JADES survey (R-F444W, G-F200W, B-F090W). We indicate the spatial scale of the images with a white bar; e) \jwst/NIRSpec R1000 spectrum from the three dispersers (G140M/F070LP, G235M/F170LP and G395M/F290LP) as part of the JADES MSA observations. The data (black line) were fitted with pPXF (see \ref{sec:jwst_data}) and the best fit (orange solid line) along with the best-fit continuum (red dashed line) are shown. The vertical-coloured dashed lines show the detection of the brightest absorption and emission lines.}
    \label{fig:data}
\end{figure}

We used the Atacama Large Millimetre Array (ALMA) to observe CO(3-2) emission line, a common tracer of molecular gas in high-redshift galaxies. We present the continuum (dust) emission, CO(3-2) map and the extracted ALMA band-3 spectrum in Fig.~\ref{fig:data}, panels a)-c). The beam size matches (0.4") the size of the galaxy. With 7-hours integration, we place a stringent 3-sigma upper limit on the integrated flux (I$_{\rm CO}$(3-2) < 0.033 Jy km~s$^{-1}$). This flux indicates a molecular cool-gas mass ranging between M$_{\rm mol}<10^{9.1}\,M_\odot$, for fiducial calibrations of CO luminosity, and $<10^{9.8}$ $\Msun$, for conservative calibrations. This value makes GS-10578 the earliest unambiguous evidence of a massive galaxy that is essentially devoid of a cool-gas reservoir, having a stellar-to-gas mass ratio of <0.8 \%. This is the most stringent upper limit on the gas mass for a quiescent galaxy at high redshift. 

\begin{figure}
        \centering
	\includegraphics[width=0.9\columnwidth]{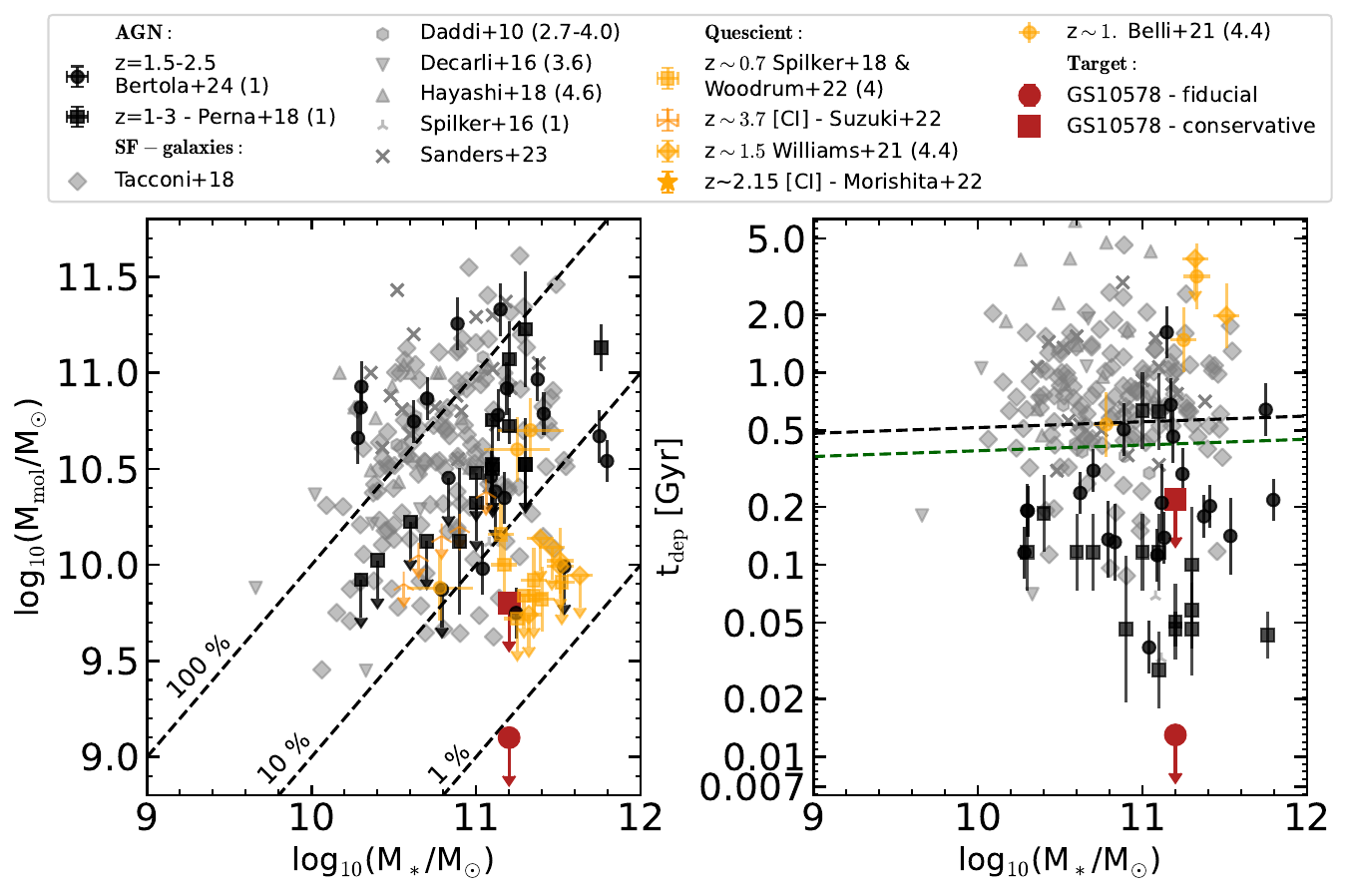}
    \caption{Molecular gas measurement of our target. Left panel: Comparison of molecular gas reservoirs of AGN, star-forming galaxies and our target as a function of stellar mass at Cosmic Noon. The dashed lines mark constant gas fractions (M$_{\rm mol }$/M$_{*}$). Right panel: Comparison of depletion time scales for AGN, star-forming galaxies and our target as a function of stellar mass. The green and black dashed lines show depletion time scales at z=3 and z=2 from \citep{Tacconi20}, respectively. The grey points indicate various star-forming galaxy samples in the redshift range of 1--3 \citep{Daddi10,Decarli16, Spilker16, Hayashi18, Tacconi18,Morishita22,Suzuki22,Sanders23}, the black points show AGN at Cosmic Noon \citep[][]{Perna18, Bertola+2024}. The yellow-coloured points show quiescent galaxies from \citep{Spilker18,Bezanson19, Belli+21,Williams21,Woodrum22}. In the legend we include $\alpha_{\rm CO}$ conversion used in each study. Our target GS10578 is shown as a red circle and a square for optimistic and conservative $\alpha_{\rm CO}$ conversion, respectively. Our deep ALMA observations provide tight constraints on the molecular gas mass at these high redshifts. Our depletion timescales are estimated based on the outflow rates rather than SFR as is for the rest of the sample.}
    \label{fig:Mgas}
\end{figure}

In Fig.~\ref{fig:Mgas}  we investigate the cool molecular gas mass (left panel) as a function of the stellar mass and compare GS-10578 to other AGN hosts (black symbols; \citep{Perna18, Bertola+2024}), quiescent galaxies (orange symbols; \citep{Belli+21,Williams21, Woodrum22}) and star-forming galaxies (various grey symbols; \citep{Daddi10, Decarli16, Spilker16, Tacconi18, Sanders23}) at z=1-4. Our ultra-deep ALMA observations allow us to place an upper limit on molecular gas mass, on par with previous studies of quiescent galaxies at z=0.7-1.5 \citep{Williams21, Spilker18}, yet 1.5~Gyr earlier in cosmic time and a factor of 10 deeper than previous studies at z$>$3 \citep{Suzuki22}, making this a unique data set. Furthermore, the gas fraction of our target is factor 3--10 lower than for other AGN at Cosmic Noon likely associated with star-forming hosts.

We used new JWST/NIRSpec multi-object spectroscopic observations from the JADES survey (see Fig. \ref{fig:data}, panels d) and e) to measure the mass-outflow rate of the neutral gas, giving $\dot{M}_\mathrm{out} =60\pm20$ \Msun yr$^{-1}$ (see Fig.~\ref{fig:nai}), in agreement with the literature \citep{DEugenio23a}. Furthermore, with this new data, we now have wavelength coverage of Pa$\beta$, although not detected. The lack of the detection of Pa$\beta$ emission line sets an upper limit on SFR of $<5.6$ \Msun yr$^{-1}$, even when we assume a conservative dust extinction of  A$_{\rm V}$ = 2. 
The neutral gas outflow dominates the mass loading factor as the mass outflow rate is a factor >20 larger than the SFR and a factor of >20 larger than the ionised gas outflow. As the neutral gas outflow is dominating the consumption/removal of the gas in this galaxy, we estimated the depletion timescale, i.e. the time necessary to consume/remove any available gas through star formation or outflows (t$_{\rm dep}$) as $<16$ or $<220$ Myr, depending on the fiducial or conservative value of M$_{\rm mol}$. This timescale is extremely rapid when compared e.g. to the main-sequence timescale (i.e. the time the galaxy spends on the star-forming main sequence), which at $z=3$ is of order 1/sSFR = 1.0~Gyr (where sSFR is the SFR per unit stellar mass) \citep{Schreiber15, Carnall23b, Carnall_spec}.


In the right panel of Fig. \ref{fig:Mgas}, we compare the t$_{\rm dep}$ of our target (red points) to those of star-forming (various grey markers) and AGN-host galaxies (black points). As we do not have any outflow properties of the comparison sample, we estimate t$_{\rm dep}$ of the comparison sample as M$_{\rm mol}$/SFR. In star-forming galaxies, the total mass loading factor ($\dot{M}_{\rm out}$/SFR) is in general $\sim1$ \citep{Fluetsch19,ForsterSchreiber19}. Therefore, when taking into account both SFR and outflows, the gas depletion time of star-forming galaxies might be $\sim0.3$ dex lower than that estimated from M$_{\rm mol}$/SFR. For AGN-host galaxies, the gas depletion is dominated by the outflows \citep[($\dot{M}_{\rm out}$/SFR)>1;][]{Fluetsch19,ForsterSchreiber19,Kakkad20}. For the majority of quiescent galaxies, we cannot estimate t$_{\rm dep}$, because these samples have upper limits on both the SFR and M$_{\rm mol}$ (except for five objects, orange points in the right panel). In these five quiescent cases, the depletion times are a factor 5-10 larger than the gas depletion times of galaxies on the main sequence with no evidence of outflows, indicating that any surviving gas is long-lived, perhaps stabilised against star formation by dynamical effects (\citep{martig09,davis11} 1/sSFR of 1 Gyr at z=1--1.5). In contrast, the depletion time of GS10578 is 3--50 times faster, indicating that gas in this early quiescent galaxy is being rapidly removed, and pointing to a profound difference with respect to quiescent galaxies at later epochs. The depletion times of our target are on par with other AGN (for the conservative upper limit) and well below for our fiducial value. We thus present the first compelling upper limit on rapid gas depletion time in a high-redshift quiescent galaxy.



\begin{figure}
        \centering	
        \includegraphics[width=0.99\columnwidth]{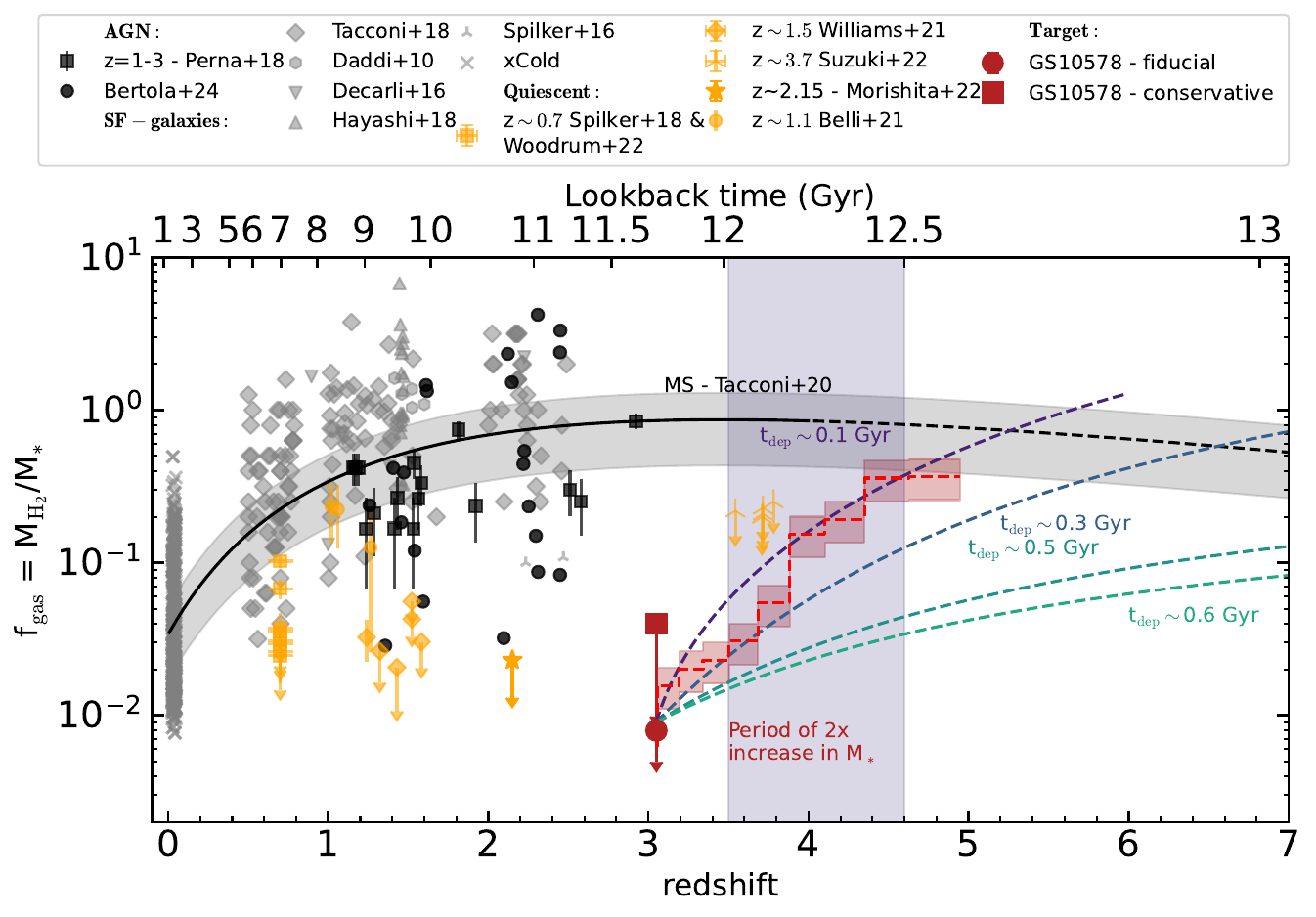}
    \caption{Molecular gas fraction as a function of redshift for star-forming, AGN hosts and quiescent galaxies. The black line indicates the gas fraction of the star-forming main sequence from \citep{Tacconi20} (with stellar mass 10$^{11}$M$_{\odot}$, i.e. mass of the galaxy at the start of the rapid growth episode) with the shaded region indicating 0.3 dex scatter of the relation. The line becomes dashed when we start extrapolating the relationship from \citep{Tacconi20}. The star-forming, AGN and quiescent galaxy comparison samples are as shown in Fig. \ref{fig:Mgas}. We also added data from local star-forming galaxies (xCold GASS - grey crosses; \citep{xCold17}). The red dashed line shows the evolution of gas fraction based on the star-formation-history from \citep{DEugenio23a} and assuming no gas inflows. 
    We highlight multiple different coloured dashed lines indicating gas fraction evolution assuming constant depletion times (t$_{\rm dep}$). The dark blue shaded region indicates time when the stellar mass experienced a rapid growth phase \citep{DEugenio23a}. The conclusion of this work do not change if we use the fiducial or conservative gas fraction data point.}
    \label{fig:fgas}
\end{figure}

Based on the unprecedented combination of deep ALMA and \jwst/NIRSpec data, we can now infer both the past and future evolution of this galaxy. The star-formation history (SFH) and molecular gas reservoir suggest that this galaxy went through rapid quenching events in the past. However, the presence of a rotating stellar disc \citep{DEugenio23a} means that the galaxy was quenched without destroying the stellar disc. This latter result implies that the current stellar mass did not form in many different halos which later merged, as the multiple merger events would destroy the stellar disc seen at the epoch of observation.


There are multiple different pathways to quench a galaxy. Quenching could be caused by ejective feedback \citep{Tremonti07, Davis12, Alatalo15,King15, Maltby19, Baron20,Belli23, Davies24}, removing the available gas from the galaxy, or through prevention of accretion of fresh gas from the intergalactic and circum-galactic medium onto the galaxy \citep{Dekel06, Dave12, Koudmani22,Scholtz23QSO}, with star formation consuming the cool gas already present in the disc (this scenario is also referred to as `starvation').


Tracing the history of the gas fraction with redshift has a great potential to distinguish between the different modes of feedback that this galaxy underwent in the past. The star formation histories were estimated by \citep{DEugenio23a}. Given the exquisite quality of the JWST data, we assume a non-parametric SFH. This setup is described in \citep{Leja18}.
They implement a `continuity' probability prior, which assumes the SFR does not change between adjacent bins, while the so-called `fat tails' of the Student's t distribution do not (over) penalize changes where they are required by the data.
The estimated SFR and stellar masses show that the galaxy was on the main sequence at $z\sim4.7$, as the galaxy's SFR was $\sim$258 \Msun/yr (compared to the expected SFR of SF galaxy of $\sim$190 \Msun/yr at redshift 4.7; \cite{Popesso23}).
We start by assuming there was no gas inflow onto the galaxy since z$\sim$4.7 (i.e. the start of the rapid growth phase), we integrated the SFH in the past 1.5 billion years, assuming a 70~per cent conversion of gas into stars (i.e. 30\% of gas is returned back to the ISM through supernovae and stellar winds \citep{Chabrier03}). We plot this evolutionary track as a red dashed line in Fig. \ref{fig:fgas}, and where the red shaded region to indicates the uncertainties. Our track estimated from the SFH of the galaxy implies an extremely short time-averaged depletion time of 0.1 Gyr, in agreement with the instantaneous estimate of 16-200 Myr derived from the
neutral-gas outflow. This agreement implies that star formation and
ejective feedback had similarly rapid timescales for this galaxy.

The SFH implies that GS-10578 started on the main sequence gas-fraction relation at $z\approx4.7$ (within uncertainties), regardless of whether we use a value of 3\% or 0.8 \% for the gas fraction. We stress that this result is not by construction. The gas-consumption history could in principle end below the star-forming main sequence, for example, if ejective feedback removed a substantial fraction of the gas, or even above the star-forming main sequence if this galaxy had higher-than-average gas inflows. The match between the gas consumption history and star-forming main sequence implies that GS10578 evolved with net-zero gas inflow , i.e. either no gas was accreted from the CGM/IGM or any accreted gas was expelled by the AGN, with no need for rapid ejective feedback episodes of the entire galaxy gas supply.

Within the large uncertainties on the SFH and on the conversion between SFH and $f_\mathrm{gas}$, we can conclude that gas consumption due to star formation was sufficient to use up most of the cool gas, hence moving this galaxy away from the high gas fractions typical of star-forming galaxies at $z=3$, and into a gas-poor `graveyard' where no appreciable star formation is possible. 

However, this scenario relies on the assumption that the galaxy halo completely stopped gas accretion; any failure or delay in preventative feedback or the impact of its complex environment would result in an influx of cool gas, which by redshift $z\approx 3.5$ needs to be either removed or otherwise prevented from forming stars (or the galaxy would not be quiescent at $z=3$). 

However, the net zero inflow is not caused by the complete halt of gas accretion on to the galaxy. We observe gaps in the preventative feedback in the form of multiple gas-laden satellites bringing gas in the innermost regions of the GS-10578 halo \citep{DEugenio23a} (as well as candidate AGN within 30 kpc, see \citep{Perna23b}). Therefore, this incomplete halt of gas accretion requires some form of mild ejective feedback at the same rate as gas inflow into the galaxy. Furthermore, we also see ejective feedback in action, as supported both by our analysis of neutral gas outflow in the MSA data and by \citep{DEugenio23a} using the IFU data and their methods. Moreover, neutral-gas outflows seem to be widespread in massive, quiescent galaxies at $z\sim2-3.7$ \citep{Belli23,Davies24, Perez-Gonzalez24}; this suggests that the conditions found in GS10578, far from being extreme, may instead be typical in early quiescent galaxies.

Our conclusions about net zero inflow scenario holds even when we assume the most conservative upper limit on the current gas mass, as the 2.7\% change in the starting point of the tracks does not influence our conclusions. Our conclusions support that although ejective feedback is in action, it does not eject the majority of the galaxy's gas supply to quench it, but only any incoming fresh gas, or heats the halo, preventing further gas accretion \citep{Koudmani22}.
This is consistent with the results of \citep{Baker23}, who found that metallicities of quiescent galaxies are consistent with the scenario of black hole quenching galaxies via halo heating (either preventative or delayed feedback), which results in lack of fresh gas accretion and lack of dilution of the metal content of a galaxy with pristine gas from the IGM. 


The existence of old, quiescent galaxies at z=0, whose formation redshifts exceed z>2  \citep{McDermid15}, suggests that galaxies like GS10578 may indeed stay quiescent for billions of years, between z=3 and z=0, perhaps growing via gas-poor mergers \citep{Perez-Gonzalez08, vanderwel14, bezanson18}.
This raises the question of what could maintain this galaxy quenched for the next 11.7 billion years until z=0. We attempt to answer this question by investigating the impact of the current AGN episode on the host galaxy. This AGN episode drives the current massive neutral and fast ionised gas outflows, removing any available gas. We attempted to quantify the age of this AGN, by investigating the velocity and extent of the ionised gas outflows using the IFS data from  \citep{DEugenio23a}. We can make a simplistic scenario that assumes that the AGN outflow velocity has not changed during the AGN episode and estimate the time it takes for the outflow to propagate to its maximum radius. Given the radius of the ionised gas outflow is 4.5 kpc (as traced by the \OIIIL) and the average ionised gas outflow velocity of 1,500 km s $^{-1}$, we estimated that it took the outflow $\sim$3 Myr to propagate to 4.5 kpc - making the AGN episode $\sim$3 Myr long. However, we stress that these are only order-of-magnitude estimates, which ignore AGN flickering on scales of $10^{4}-10^{5}$ years and varying outflow velocity
\citep{Schawinski15, Decker23}.

Knowing the age of the AGN episode, we can attempt to estimate the amount of gas removed from the galaxy by the outflows. We assume that the outflow rate in the \NaI has not changed and lasted for full $\sim3\times 10^{6}$ years (which is an optimistic scenario). Therefore, if we stack these assumptions in favour of the AGN to estimate the most optimistic impact of this powerful AGN, the AGN removed at most $\sim 2\times 10^{8}$ M$_{\odot}$ of gas from this galaxy, $\gtrsim$15~\% of the maximum gas mass permitted by ALMA, with this long and luminous AGN episode. Therefore, a single AGN episode can possibly damage the small gas reservoirs of quiescent galaxies, however, it would not majorly influence the evolution of gas-rich star-forming galaxies, requiring cumulative feedback from all the AGN episodes 
\citep{Harrison18,Scholtz20, Scholtz21, Lamperti21,Piotrowska22, Baker23, Harrison24}. 

The old age of the stellar populations clearly means that GS-10578 became quiescent hundreds of Myr before the current AGN episode. Given the low gas content of this galaxy, the current AGN episode could have been triggered by the interaction with the low-mass star-forming companions \citep{DEugenio23a,Perna23b}. Overall, our stringent upper limit on the molecular gas mass allows us to observe the preventative mode of AGN feedback in these quiescent galaxies. Future work needs to focus on the gas reservoirs around these galaxies; preventative vs ejective feedback leaves distinctive signatures on the hot gas, which may be probed by sub-mm observations \citep{Brownson19,Jones22} or by future X-ray observatories.

Still, our work demonstrates the power of combining accurate SFH from JWST and stringent measurements of the gas fraction from ALMA. Based on these new ultra-deep ALMA observations, we showed that the gas fraction and depletion times of early quiescent galaxies could be lower and faster than in quiescent galaxies at z=1.5, possibly due to more frequent or more efficient AGN. However, this needs to be confirmed by a large sample of high-z early quiescent galaxies with deep gas fraction constraints. Comparing the gas-consumption history with the star-forming main sequence, we are able to investigate the distinctive role of preventative vs ejective feedback. Unfortunately, from a single galaxy, we cannot draw strong conclusions; larger sample sizes are required.

\section{Methods}

\subsection{ALMA observations \& analysis}\label{sec:ALMA_data}

We observed the CO(3-2) emission line of GS-10578 with ALMA as part of programme 2023.1.01296.S (PI: D'Eugenio). CO(3-2) is a commonly used transition at high redshift to trace cold molecular gas due to its accessibility with e.g. ALMA and NOEMA \citep[e.g. ][]{Circosta21, Tacconi20}. The data were taken from 18\textsuperscript{th} November to 12\textsuperscript{th} of December 2023 in Band 3, with a baseline range of 15--3690 m. The total on-source
exposure was $\sim7$ hours. The raw science data models were processed by the \texttt{scriptForPI} script, which resulted in the calibrated measurement sets for each observation. As the source was observed over six separate executions, we joined these executions using \textsc{CASA}'s \texttt{concat} task. The measurement sets also included the calibrator sources,  so we created measurement sets containing only calibrated visibilities of the target source with \textsc{CASA}'s \texttt{split} task and time binned in 30s intervals for easier data management. 

The calibrated visibilities were imaged using \texttt{tclean} onto cubes using the natural weighting scheme (to maximise the sensitivity) with a final beam size of 0.64-0.69 arcsec. To maximise the sensitivity of the final emission-line cubes, they were imaged with 100 km s$^{-1}$ velocity channels, assuming that CO and ionised lines have the same FWHM. This allows us to resolve any potential emission lines with $>$3 channels across the emission line. The final emission-line cubes have RMS of 0.06 mJy/beam in the 100 km s$^{-1}$ velocity bin. We also imaged the continuum by excluding any spectral channels in which we would expect the CO(3-2) emission based on the redshift and velocity dispersion of the lines in the NIRSpec spectra. The final rms of the continuum maps is 0.004 mJy/beam. We cleaned the imaged data using \textsc{CASA}'s \texttt{tclean} task down to 3-$\sigma$ depth; however, we verified that any detections or non-detections are consistent in both the dirty (not cleaned) and cleaned imaged data. 

We show the 3-mm continuum image in the a) panel of Fig.~\ref{fig:data}, with the beam indicated as a magenta ellipse in the bottom left corner. There is no evidence for a $>3$-$\sigma$ detection within 1 arcsec of the target. We note a 3.5-$\sigma$ feature southwest of the target, but without a NIRCam or Ly\textalpha\ counterpart, and below the detection threshold for ALMA \citep{Kaasinen23}. 

Finally, we created a moment-0 map of CO(3-2) using Python's \texttt{Spectralcube} by collapsing the cleaned emission-line data cube in $\pm$150 km s$^{-1}$ around the systemic redshift of our target and we show the map in the b) panel of Fig. \ref{fig:data}. We highlight $\pm$2,3,4-$\sigma$ contours of the moment-0 image as red dashed and solid lines. Furthermore, we extracted a spectrum centred on the \jwst/NIRCam position with an aperture corresponding to the beam size and we show it in the c) panel of Fig. \ref{fig:data}. We do not see any $>3$-$\sigma$ detection in either the extracted spectrum or the moment-0 maps. We also verified our non-detection by repeating the analysis on the dirty images to confirm that this non-detection was not caused by incorrect cleaning.

\begin{table}
   \caption{Main properties and ALMA results for the main target GS-10578.}
   \centering
 \begin{tabular}{lr}
  \hline
  \hline
  RA (deg) & 53.165314\\
  Dec (deg) & -27.814136\\
  z$_{\rm spec}$ & 3.064 \\
  log$_{10}$($L_{\rm bol, AGN}$/erg s$^{-1}$) &46.7$\pm$0.5\\
  log$_{10}$(M$_{*}$/M$_{\odot}$) & 11.2$\pm$0.1\\
  log$_{10}$(M$_{\rm dyn}$/M$_{\odot}$) & 11.3$\pm$0.1\\
  SFR (M$_{\odot}$ yr$^{-1}$) & <5.6$^\ddag$\\
  $\dot{M}_{\rm out}$ (M$_{\odot}$ yr$^{-1}$)& 58$^{+25}_{-20}$\\
  \hline
  I$_{\rm CO(3-2)}$ (Jy km s$^{-1}$) & $<0.033$\\
  f$_{\rm 3 mm}$ (mJy) & <0.012\\
  $L'_{\textnormal{CO(3-2)}}$ (K km s$^{-1}$ pc$^{2}$) & $<1.6\times 10^{9}$ \\
  M$_{\rm gas}$ (M$_{\odot}$) & $<10^{9.1}$--$10^{9.8}$\\
  t$_{\rm dep}$ (Myr) &  <16-210\\
 \hline
 \end{tabular}
 
 $^\ddag$ Obtained from the measured Paschen-$\beta$ flux, assuming no AGN contribution, hence the true SFR is likely much lower.
  \label{table:target}
\end{table}

As we described above, we do not detect any CO(3-2) emission in this galaxy and we could only place a 3-$\sigma $ upper limit on the integrated flux (I$_\mathrm{CO(3-2)}$) of 0.028 Jy km s$^{-1}$. The upper limit on the integrated flux was estimated by using the RMS of the CO(3-2) cube and assuming a width of 300 km s$^{-1}$. This FWHM matches the velocity dispersion of the stellar component, which should be kinematically closer than the ionised gas kinematics (traced by \OIIIall), which is heavily influenced by the fast ionised gas outflows. We derived CO(3-2) luminosities (in Table~\ref{table:target}) as given by \citep{solomon05}:
\begin{equation}
L'_{\textnormal{CO}} = 3.25 \times 10^7 I_{\textnormal{CO}} \nu_{\textnormal{obs}}^{-2} D_{\textnormal{L}}^2 (1+z)^{-3},
\end{equation}
where $I_{\textnormal{CO}}$ is the velocity-integrated flux, $D_{\textnormal{L}}$ is the luminosity distance, $\nu_{\textnormal{obs}}$ is the observed frequency of the line (in this case CO-J=3-2: 85.08 GHz) and $z$ is the redshift of the galaxy. 
The CO(3-2) is tracing molecular gas with excitation temperature of 30-50 K with a critical density of 300 cm$^{-3}$.
We derive an $L'_{\textnormal{CO}}$ of $< 1.4\times 10^{9}$ K km s$^{-1}$ pc$^2$. To convert this CO(3-2) measurement to molecular gas mass, we need to assume $r_{31}$ = $L'_{\textnormal{CO(3-2)}}/L'_{\textnormal{CO(1-0)}}$ and $\alpha_{\textnormal{CO}}$. Given that this is a quiescent galaxy with a luminous AGN, we assume that the primary source of excitation of the CO molecule is the AGN and not star formation. As such we use $r_{31}$ = 0.97 derived for QSOs \citep{Carilli13, Kirkpatrick19}. As for $\alpha_{\textnormal{CO}}$, this can be dependant on metallicity and ISM conditions (temperature, density, etc, see \citep{Bolatto13}). We derive two gas masses using an optimistic and a conservative $\alpha_{\textnormal{CO}}$ value of 0.8 (value derived for QSOs; \citep{Kaasinen24}) and 4.4 M$_{\odot}$/(K km s$^{-1}$ pc$^2$) (value for quiescent galaxies \citep{Williams21}), respectively. This range covers most AGN hosts and quiescent galaxies and is the main source of systematic uncertainties on the M$_{\rm gas}$. With these assumptions we estimate the molecular gas reservoir of our target to be $<10^{9.1}$--$10^{9.8}$ $\Msun$.

We investigate the effect of $r_{31}$ and resolution on our upper limit estimated upper limit on the M$_{\rm gas}$. The effective radius (r$_{e}$) of the galaxy is $\sim$0.15 arcseconds and therefore, the beam with a size of 0.4 arcseconds well covers the expand of the expected CO emission. Furthermore, we reimaged the ALMA data with 0.7 arcsecond Gaussian taper which resulted in imaged data set with resolution of 1.3x1.0 arcseconds and 3$\sigma$ upper limit on $I_{\textnormal{CO}}$ of 0.38 Jy km s$^{-1}$. This would result in increase of resulting gas masses by 0.06 dex, having no impact on the conclusions of this work. Similarly, assumptions about the excitation of the molecular gas has impact on the assumed r$_{31}$ and hence the estimated M$_{\rm gas}$. Given that the AGN in the bolometric luminosity of target is 10$^{46.7}$ ergs s$^{-1}$ cm$^{-2}$  and the SFR is $<$5.6 \Msun yr$^{-1}$, the excitation of the molecular gas is dominated by the AGN rather than star-formation. Using a value of 0.56 for r$_{31}$ (assuming molecular gas is excited by star-formation) would increase the upper limit on the gas by $\sim$0.18 dex, with no impact on the conclusions of our work.

\subsection{\jwst/NIRSpec MSA observations \& Analysis}\label{sec:jwst_data}

\subsubsection{Observations}

New \jwst/NIRSpec MSA observations of GS-10578 were obtained by JADES(\jwst Advanced Deep Extragalactic Survey; \citep[][]{Eisenstein23}), using the NIRSpec micro-shutter assembly (MSA; \citep[][]{Ferruit22} in programme PID 1180 (JADES NIRCam~ID 197911). We use medium-resolution observations ($R=500-1,500$) spanning wavelengths $0.6<\lambda<5.3~\mu$m, obtained by splicing data from the three disperser/filter combinations: G140M/F070LP, G235M/F170LP and G395M/F290LP.
The full target selection and observation setup is described in \citep{DEugenio23a}.
The observations were performed using three-shutter slitlets with nodding, and consisted of three nods and two integrations per nod, totalling 6.2~ks per grating.

The \jwst/NIRSpec MSA observations for this target were processed with the data reduction pipeline of the ESA NIRSpec Science Operations Team (SOT) and the NIRSpec GTO Team, further detailed description of the GTO pipeline is summarised in \citep{DEugenio24}. Here, we briefly summarise the data reduction. We retrieve the level-1a products from the MAST archive and fit the count-rate maps to derive their slopes using only unsaturated groups. During this procedure, we also remove any jumps that appear in the data due to cosmic rays. At the same time, we perform master dark and bias subtraction as well as any flagging of bad data. The background subtraction is performed pixel by pixel, by combining and subtracting pairs of far-away nods (to avoid self subtraction). We performed the flat-field correction of the spectrograph optics and disperser corrections on the 2D-dimensional cutouts of each of the three-shutter slits. The path loss corrections are calculated assuming point sources and taking into account the source location on the shutter.

We extracted the 1D spectra from the combined 2D maps adopting a box-car 5-pixel aperture centred on the target. We combined all 1D spectra and removed bad pixels with sigma clipping.
In addition to the data reduction steps reported in \citep{DEugenio24}, we also rescale the G140M and G395M spectra to match the normalisation of the G235M spectrum; this is performed by comparing the median flux density in the overlap region between pairs of adjacent gratings (we downscale G140M by 0.892 and G395M by 1.078). The extracted spectrum is shown in Fig.~\ref{fig:data}, highlighting the simultaneous presence of a strong Balmer break and Balmer absorption lines, and broad emission lines, including a range of low- and high-ionisation species. We note that the absolute flux calibration of the spectrum is still uncertain, due to the point-source assumption in the path-loss correction and to lingering uncertainties in the flux calibration of NIRSpec itself. However, in the following, we are interested primarily in the equivalent width of \NaI absorption, which is not affected by these flux-calibration issues.

GS-10578 was also observed with NIRSpec/IFU within the GA-NIFS programme (`Galaxy Assembly with NIRSpec IFS'; PID 1216; PI: N. Lützgendorf), with the prism and high-resolution G235H grating \citep{DEugenio23a}. These deep high-resolution data allowed us to trace gas and stellar kinematics at this high redshift. Here we add the NIRSpec/MSA observations, which provide the ideal combination of a wide wavelength range and adequate spectral resolution.
In addition, when compared to the high-resolution NIRSpec/IFS observations, the MSA observations are significantly deeper at the continuum (14.4~ks with G235H/F170LP vs 7.2~ks hours with the medium-resolution gratings). Nevertheless, the superior spatial coverage of the IFS enabled \citep{DEugenio23a} to accurately measure the spatial extent of the gas absorption, which cannot be done with the MSA (e.g., \citep{Davies24})

\subsubsection{SED fitting and star-formation histories}

\subsubsection{Neutral gas outflow fitting}

We use \ppxf \citep{Cappellari2023} to model simultaneously the stellar continuum and nebular emission lines, following \citep{Looser23_SFH, DEugenio23a, DEugenio24}. The stellar continuum is a linear combination of simple stellar-population spectra with MIST isochrones \citep{choi+2016} and C3K model atmospheres \citep{conroy+2019}. These templates are complemented by a set of Gaussians to capture nebular emission, and by 10\textsuperscript{th}-order multiplicative Legendre polynomials.
Crucially, we add Gaussian gas-absorption templates to reproduce prominent ISM absorption features, including \MgIIall, \CaIIall and \NaIall.
This decision follows from the known presence of neutral-phase outflows in this galaxy \citep{DEugenio23a}; this means that the kinematics of these absorption lines do not match the stellar kinematics. For \NaIall, the observed equivalent width cannot be reproduced by stellar-population models with standard chemical abundances \citep{DEugenio23a}.
The resulting best-fit spectrum is shown in Fig.~\ref{fig:data} in panel e). We show the spectrum across all three gratings as a black line, with best-fit total model and continuum model as orange and red dashed lines, respectively. 

\begin{figure}
    \centering
	\includegraphics[width=0.9\columnwidth]{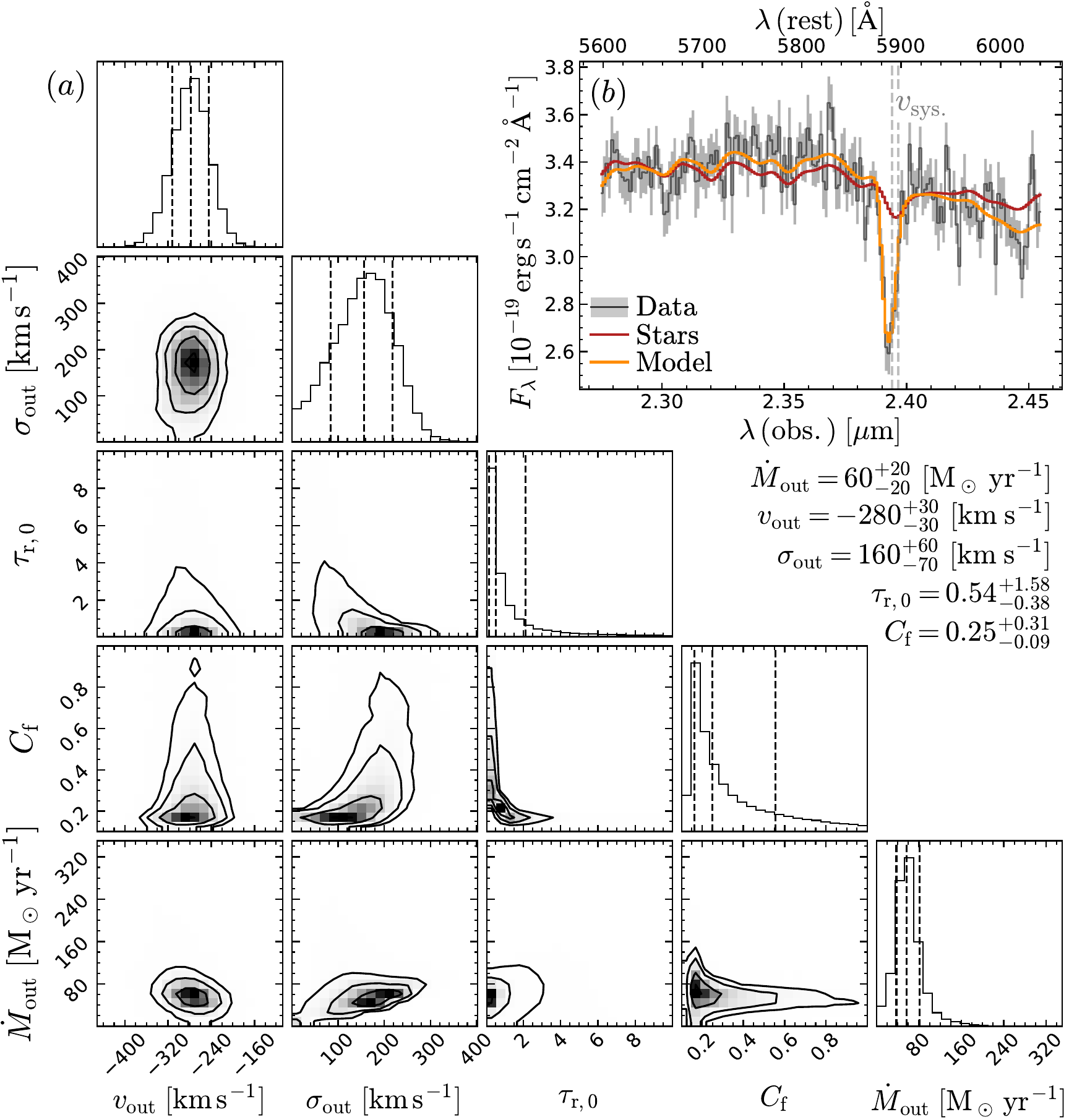}
    \caption{Fitting of the neutral gas properties. Posterior probability distribution for the neutral-gas outflow properties (panel~a) and fiducial model (panel~b) for the \NaIall absorption, which we interpret as a neutral-gas outflow. The vertical dashed lines in panel~b indicate the doublet line positions at the velocity of the stars.
    The mass outflow rate $\dot M_\mathrm{out}$ is calculated from the posterior.
    }\label{fig:nai}
\end{figure}

We estimate the mass-outflow rate from the \NaIall
absorption (following \citep[][]{rupke+2005b,Davies24}). The relevant region of the spectrum is shown in Fig.~\ref{fig:nai}; to model the stellar continuum we use the best-fit stellar templates from \ppxf and a 2\textsuperscript{nd}-order polynomial; we do not include the \HeIL[5785] line in the fit, because this line is not detected.
With this method, we estimate a mass-outflow rate of $60\pm20$ M$_\odot$~yr$^{-1}$, in agreement with estimates from the previous \jwst/NIRSpec IFS observations in the range 30--100~M$_\odot$~yr$^{-1}$; \citep{DEugenio23a}. \citep{DEugenio23a} also measured the outflow rate in the ionised gas phase using \OIIIall, finding an estimated outflow rate of 0.14--2.9 \Msun yr$^{-1}$. Due to ionised gas phase being negligible compared to the neutral gas outflow phase, we will use only the outflow rate from the neutral gas from now on.  

It is interesting to point out that despite the small amount of molecular gas reservoir, there seems to be a sufficient amount of neutral gas in order to detect the neutral gas outflow in the \NaI absorption, suggesting a large reservoir of neutral gas. During the estimation of the neutral gas outflow, we assume that 90~per cent of Na is ionised \citep{rupke+2005b}. Yet, given the lack of molecular gas, this assumption is not necessarily true. However, increasing this fraction to e.g., close to 99~per cent would boost the outflow rate to $\sim$300 \Msun yr$^{-1}$. This would decrease the depletion time scale by a factor of 3. However, we choose the conservative value of 90\% in our depletion timescales.

\bigskip

\noindent{}
\FloatBarrier
\textbf{Data availability}
The JWST data that support the findings of this study are publicly available from the 
\href{Mikulski Archive for Space Telescopes}{https://archive.stsci.edu/}. This paper makes use of the following ALMA data: ADS/JAO.ALMA\#2023.1.01296.S, available from the \href{ALMA archive}{https://almascience.eso.org/aq/}

\noindent{}
\textbf{Code availability}
This work made extensive use of the freely available
\href{Python}{http://www.python.org} programming language
\citep{vanrossum1995}, maintained and distributed by the Python Software
Foundation. We further acknowledge direct use of
\href{astropy}{https://pypi.org/project/astropy/} \citep{astropyco+2013},
\href{dynesty}{https://pypi.org/project/dynesty/} \citep{speagle2020},
\href{fsps}{https://github.com/cconroy20/fsps} \citep{conroy+2009, conroy_gunn_2010},
\href{matplotlib}{https://pypi.org/project/matplotlib/} \citep{hunter2007},
\href{numpy}{https://pypi.org/project/numpy/} \citep{harris+2020},
\href{prospector}{https://github.com/bd-j/prospector} \citep{johnson+2021}
\href{python-fsps}{https://pypi.org/project/dynesty/} \citep{johnson_pyfsps_2023},
\href{qfitsview}{https://www.mpe.mpg.de/~ott/QFitsView/},
and \href{scipy}{https://pypi.org/project/scipy/} \citep{jones+2001}.
\href{Spectral-Cube}{https://github.com/radio-astro-tools/spectral-cube}
\backmatter

\section*{Acknowledgements}
{\small 
We thank Jake Bennett, Martin Bourne, Sophie Koudmani and Debora Sijacki for a fruitful discussion that helped improve this work. This paper makes use of the following ALMA data: ADS/JAO.ALMA\#2023.1.01296.S. ALMA is a partnership of ESO (representing its member states), NSF (USA) and NINS (Japan), together with NRC (Canada), NSTC and ASIAA (Taiwan), and KASI (Republic of Korea), in cooperation with the Republic of Chile. The Joint ALMA Observatory is operated by ESO, AUI/NRAO and NAOJ.
JS, FDE, RM, WMB, TJL and JW acknowledge support by the Science and Technology Facilities Council (STFC), ERC Advanced Grant 695671 ``QUENCH'' and the
UKRI Frontier Research grant RISEandFALL.
RM also acknowledges funding from a research professorship from the Royal Society. PGP-G acknowledges support from grant PID2022-139567NB-I00 funded by Spanish Ministerio de Ciencia e Innovaci\'on MCIN/AEI/10.13039/501100011033,
FEDER {\it Una manera de hacer
Europa}.
SC, EP and GV acknowledge support from the European Union (ERC, WINGS, 101040227). 
IL acknowledges support from PID2022-140483NB-C22 funded by AEI 10.13039/501100011033 and BDC 20221289 funded by MCIN by the Recovery, Transformation and Resilience Plan from the Spanish State, and by NextGenerationEU from the European Union through the Recovery and Resilience Facility.
MP, SA and BRdP acknowledge grant PID2021-127718NB-I00 funded by the Spanish Ministry of Science and Innovation/State Agency of Research (MICIN/AEI/ 10.13039/501100011033); MP also acknowledges the grant RYC2023-044853-I, funded by  MICIU/AEI/10.13039/501100011033 and European Social Fund Plus (FSE+). 
The work of CCW is supported by NOIRLab, which is managed by the Association of Universities for Research in Astronomy (AURA) under a cooperative agreement with the National Science Foundation. 
AJB and GCJ acknowledge funding from the ``FirstGalaxies'' Advanced Grant from the European Research Council (ERC) under the European Union’s Horizon 2020 research and innovation programme (Grant Agreement No. 789056).
GC and EB acknowledge the support of the INAF Large Grant 2022 ``The metal circle: a new sharp view of the baryon
cycle up to Cosmic Dawn with the latest generation IFU facilities''
SA acknowledges support from the JWST Mid-Infrared Instrument (MIRI) Science Team Lead, grant 80NSSC18K0555, from NASA Goddard Space Flight Center to the University of Arizona.
BER acknowledges support from the NIRCam Science Team contract to the University of Arizona, NAS5-02015, and JWST Program 3215. The authors acknowledge use of the lux supercomputer at UC Santa Cruz, funded by NSF MRI grant AST 1828315.
H{\"U} gratefully acknowledges support by the Isaac Newton Trust and by the Kavli Foundation through a Newton-Kavli Junior Fellowship. The project leading to this publication has received support from ORP, that is funded by the European Union’s Horizon 2020 research and innovation programme under grant agreement No 101004719 [ORP].
}

\section*{Author Contributions Statement}
{\small 
J.S. reduced the ALMA data and measured gas content, wrote the main text, and created the figures. F.D'E designed the ALMA observations and performed the outflow calculations and wrote the main text. F.D'E. and  P.G.P.-G. performed the SED fitting. R.M. designed the survey and observations and helped with the interpretation of the results.  S.C. reduced the MSA data. B.R. and S.T. provided the imaging data. C.J.W. helped with data reduction. In addition, all authors helped at various stages of the survey design and execution, read the manuscript, and provided comments and insight for the interpretation.}

\section*{Competing interests}
{\small 
The authors declare no competing interests.}

\newcommand\aap{A\&A}                
\let\astap=\aap                          
\newcommand\aapr{A\&ARv}             
\newcommand\aaps{A\&AS}              
\newcommand\actaa{Acta Astron.}      
\newcommand\afz{Afz}                 
\newcommand\aj{AJ}                   
\newcommand\ao{Appl. Opt.}           
\let\applopt=\ao                         
\newcommand\aplett{Astrophys.~Lett.} 
\newcommand\apj{ApJ}                 
\newcommand\apjl{ApJ}                
\let\apjlett=\apjl                       
\newcommand\apjs{ApJS}               
\let\apjsupp=\apjs                       
\newcommand\apss{Ap\&SS}             
\newcommand\araa{ARA\&A}             
\newcommand\arep{Astron. Rep.}       
\newcommand\aspc{ASP Conf. Ser.}     
\newcommand\azh{Azh}                 
\newcommand\baas{BAAS}               
\newcommand\bac{Bull. Astron. Inst. Czechoslovakia} 
\newcommand\bain{Bull. Astron. Inst. Netherlands} 
\newcommand\caa{Chinese Astron. Astrophys.} 
\newcommand\cjaa{Chinese J.~Astron. Astrophys.} 
\newcommand\fcp{Fundamentals Cosmic Phys.}  
\newcommand\gca{Geochimica Cosmochimica Acta}   
\newcommand\grl{Geophys. Res. Lett.} 
\newcommand\iaucirc{IAU~Circ.}       
\newcommand\icarus{Icarus}           
\newcommand\japa{J.~Astrophys. Astron.} 
\newcommand\jcap{J.~Cosmology Astropart. Phys.} 
\newcommand\jcp{J.~Chem.~Phys.}      
\newcommand\jgr{J.~Geophys.~Res.}    
\newcommand\jqsrt{J.~Quant. Spectrosc. Radiative Transfer} 
\newcommand\jrasc{J.~R.~Astron. Soc. Canada} 
\newcommand\memras{Mem.~RAS}         
\newcommand\memsai{Mem. Soc. Astron. Italiana} 
\newcommand\mnassa{MNASSA}           
\newcommand\mnras{MNRAS}             
\newcommand\na{New~Astron.}          
\newcommand\nar{New~Astron.~Rev.}    
\newcommand\nat{Nature}              
\newcommand\nphysa{Nuclear Phys.~A}  
\newcommand\pra{Phys. Rev.~A}        
\newcommand\prb{Phys. Rev.~B}        
\newcommand\prc{Phys. Rev.~C}        
\newcommand\prd{Phys. Rev.~D}        
\newcommand\pre{Phys. Rev.~E}        
\newcommand\prl{Phys. Rev.~Lett.}    
\newcommand\pasa{Publ. Astron. Soc. Australia}  
\newcommand\pasp{PASP}               
\newcommand\pasj{PASJ}               
\newcommand\physrep{Phys.~Rep.}      
\newcommand\physscr{Phys.~Scr.}      
\newcommand\planss{Planet. Space~Sci.} 
\newcommand\procspie{Proc.~SPIE}     
\newcommand\rmxaa{Rev. Mex. Astron. Astrofis.} 
\newcommand\qjras{QJRAS}             
\newcommand\sci{Science}             
\newcommand\skytel{Sky \& Telesc.}   
\newcommand\solphys{Sol.~Phys.}      
\newcommand\sovast{Soviet~Ast.}      
\newcommand\ssr{Space Sci. Rev.}     
\newcommand\zap{Z.~Astrophys.}       
\bibliographystyle{ancillary/sn-matphys}
\bibliography{ancillary/astrobib}

\end{document}